\documentclass{elsart}

\usepackage{epsfig}
\usepackage{amssymb,amsmath}

\usepackage{url}

\begin{document}

\rightline{DO-TH 07/01}
\rightline{LPHE/07-01}

\begin{frontmatter}



\title{Photon polarization from helicity suppression in
radiative decays of polarized $\boldsymbol{\Lambda_b}$ to spin-3/2 baryons}


\author[Dortmund]{G.~Hiller}, 
\author[EPFL]{M.~Knecht}, 
\author[EPFL]{F.~Legger\thanksref{Munich}}, and 
\author[PSI]{T.~Schietinger}
\address[Dortmund]{Institut f\"ur Physik, Universit\"at Dortmund, D-44221 Dortmund, Germany}
\address[EPFL]{Laboratory for High-Energy Physics, Ecole Polytechnique F\'ed\'erale, \\ CH-1015 Lausanne, Switzerland}
\address[PSI]{Paul Scherrer Institut, CH-5232 Villigen PSI, Switzerland}

\thanks[Munich]{Now at Max-Planck Institute for Physics, D-80805 M\"unchen, Germany}

\vspace{1cm}

\begin{abstract}
We give a general parameterization of the $\Lambda_b \rightarrow
\Lambda(1520)\gamma$ decay amplitude, applicable to any strange isosinglet
spin-3/2 baryon, and calculate the branching fraction and helicity amplitudes.
Large-energy form factor relations are worked out, and it is shown 
that the helicity-3/2 amplitudes vanish at lowest order in soft-collinear 
effective theory (SCET).
The suppression can be tested experimentally at the LHC and elsewhere, thus
providing a benchmark for SCET.
We apply the results to assess the experimental reach for a possible
wrong-helicity $b \to s \gamma $ dipole coupling in 
$\Lambda_b \rightarrow \Lambda(1520)\gamma \to p K \gamma$ decays.
Furthermore we revisit $\Lambda_b$-polarization at hadron colliders and update the
prediction from heavy-quark effective theory.
Opportunities associated with $b \rightarrow d \gamma$ afforded by high-statistics 
$\Lambda_b$ samples are briefly discussed in the general context of CP and 
flavour violation.
\end{abstract}

\end{frontmatter}

% main text
\newpage
%%%%%%%%%%%%%%%%%%%%%%%%%%%%%%%%%%%%%%%%%%%%%%%%%%%%%%%%%%%%%%%%%%%%%%%%%%%%%%%%%%%%%%%%%%%
\section{Introduction}
%%%%%%%%%%%%%%%%%%%%%%%%%%%%%%%%%%%%%%%%%%%%%%%%%%%%%%%%%%%%%%%%%%%%%%%%%%%%%%%%%%%%%%%%%%%

Flavour-changing neutral current decays of the $b$-quark, such as 
$b \to s \gamma$ and $b \to d \gamma$, are important 
probes of the flavour sector of the theory \cite{SuperB}.
While the rates of e.g.\ $B \to X_s \gamma$ and $B \to K^* \gamma$
are in agreement with the Standard Model (SM), the handedness
of the underlying electromagnetic dipole transition $b \to s \gamma$, which 
in the SM requires predominantly left-polarized photons \cite{Inami-Lim},
is poorly constrained to date \cite{Schietinger}.

Radiative decays of $b$-flavoured baryons allow the study of spin correlations, 
giving information on the chirality of the dipole 
transition. 
The decay $\Lambda_b \to \Lambda(1116) \gamma$
with subsequent dominant $\Lambda(1116)$ decay to $p \pi^-$ has 
been identified as a useful mode
to test the SM at colliders \cite{GKS-MR,Hiller-Kagan,Zhao}.
In particular, the spin of the $\Lambda(1116)$ is self-analyzed by its 
decay, i.e., correlated with the direction of the momentum of the proton.
A second spin-asymmetry can be formed if the $\Lambda_b$'s are polarized 
\cite{Hiller-Kagan}.
The main drawback of experimental studies involving $\Lambda(1116)$, however,
is its weak decay:
Due to the  associated long decay length, a large fraction of these baryons
decays outside the inner (vertex) part of a given detector, posing formidable 
difficulties for the decay reconstruction \cite{Legger-Schietinger}.

For this reason, it was suggested in Ref.~\cite{Legger-Schietinger} to study
$b \to s \gamma$-mediated $\Lambda_b$-decays to heavier $\Lambda$-baryons
(resonances), which decay 
strongly and copiously into $N \overline{K}$ \cite{PDG06}.
The self-analyzing property of the $\Lambda$-spin is lost for these
modes, and photon helicity extraction therefore requires
known and finite $\Lambda_b$-polarization. 
For $\Lambda_b$'s originating from energetic $b$-quarks
heavy-quark effective theory predicts a large fraction
of the longitudinal $b$-quark polarization to be retained after hadronization
\cite{Mannel-Schuler,Falk-Peskin}.
In fact, $\Lambda_b$-polarization is found to be substantial in 
$e^+ e^- \to Z \to b \bar b$ reactions
in agreement with the $Z b \bar b$-couplings of the SM \cite{ALEPH-OPAL}.
There is no data yet on the polarization of $\Lambda_b$'s produced in a hadronic
environment.
Expectations based on perturbative QCD combined with recent experimental data 
yield a polarization not exceeding the 10\% level in high-energy $pp$ collisions,
see Sec.~\ref{sec:pol}.
While the anomalously large polarization observed in $\Lambda$ production still 
lacks theoretical understanding, it could suggest larger polarization also
for $\Lambda_b$ production. 

The theoretical framework for radiative $\Lambda_b$ decays to 
$J=1/2$ $\Lambda$-baryons can be inferred from
corresponding work on $\Lambda_b \to \Lambda(1116) \gamma$ decays,  
e.g., Ref.~\cite{Hiller-Kagan}.
There is, however, no calculation available for $J=3/2$ (or higher).
In particular, it is not known whether, for a given photon handedness,
the decay amplitudes to the $\pm$1/2 and $\pm$3/2 helicity states differ 
significantly, as required for the extraction of the photon helicity 
along the lines of Ref.~\cite{Legger-Schietinger}.
It is the purpose of this paper to fill this gap.
Specifically, we will work out helicity amplitudes and rates 
for $\Lambda_b \to \Lambda(1520) \gamma $ decays.  
The $\Lambda(1520)$ with  $J^P=3/2^-$ is expected to produce  
a prominent peak in the $p K$-mass spectrum from 
$\Lambda_b \to p K  \gamma$ decays due to its
large branching fraction to $p K$ and its relatively narrow width 
\cite{Legger-Schietinger}.
Soft-collinear effective theory (SCET) simplifies strong-interaction effects 
such as form factors in exclusive heavy-to-light decays at large recoil
\cite{CLOPR,BFPS}.
We work out lowest-order form factor relations in $\Lambda_b \to
\Lambda(1520) \gamma $ decays and show that the amplitude with $\Lambda$-helicity 
$\pm 1/2$ dominates, thus supporting the experimental 
extraction of the photon helicity in $\Lambda_b \to \Lambda (J=3/2) \gamma \to 
p K \gamma$ modes according to Ref.~\cite{Legger-Schietinger}.
Furthermore, we point out that measurements of $\Lambda_b \to \Lambda(1520)  
\gamma \to p K \gamma$ angular distributions afford a quantitative test of
the suppression of the helicity-3/2 amplitude.

%%%%%%%%%%%%%%%%%%%%%%%%%%%%%%%%%%%%%%%%%%%%%%%%%%%%%%%%%%%%%%%%%%%%%%%%%%%%%%%%%%%%%%%%%%%
\section{$\boldsymbol{\Lambda_b \to \Lambda(1520) \gamma}$ amplitude and branching fraction}
%%%%%%%%%%%%%%%%%%%%%%%%%%%%%%%%%%%%%%%%%%%%%%%%%%%%%%%%%%%%%%%%%%%%%%%%%%%%%%%%%%%%%%%%%%%
\label{sec:amplitude-and-bf}

In the following, we use $\Lambda$ to denote the $\Lambda(1520)$-baryon 
unless otherwise stated.
The description of weak decay amplitudes is done in an effective 
low-energy theory framework \cite{Buchalla} starting from the effective
Hamiltonian
\begin{equation}
{\mathcal{H}}_{\rm eff}= -4 \frac{G_F}{\sqrt{2}} 
V_{tb} V_{ts}^* \sum c_i {\mathcal{O}}_i + c_i^\prime {\mathcal{O}}_i^\prime+ {\mathrm h.c.}\, ,
\end{equation}
where the chirality-flipped operators  ${\mathcal{O}}_i^\prime$ are obtained from
the V--A-operators ${\mathcal{O}}_i$ by interchanging the chiral projectors  
$L/R=(1 \mp \gamma_5)/2$. 
The most important contributions are due to the electromagnetic
dipole operators 
\begin{equation}
{\mathcal{O}}_7 = \frac{e}{16 \pi^2} m_b \bar s \sigma_{\mu \nu} F^{\mu \nu} R b \, , \qquad
{\mathcal{O}}_7^\prime = \frac{e}{16 \pi^2} m_b \bar s \sigma_{\mu \nu} F^{\mu \nu} L b \, ,
\end{equation}
where $F^{\mu \nu}$ denotes 
the electromagnetic field strength tensor. Within the
SM, the flipped dipole operator ${\mathcal{O}}_7^\prime$ is suppressed 
by the quark mass ratio $m_s/m_b$ with respect to  ${\mathcal{O}}_7$, 
leading to predominantly left-handed photons in $b \to s \gamma$ quark decays.

The $\Lambda_b \to \Lambda(1520) \gamma$ amplitude can then
be written as
\begin{align}
i {\mathcal{M}}_{fi} &= 
\langle \gamma (q, \epsilon) \Lambda (p^\prime, s^\prime) |{\mathcal{H}}_{\rm eff} | 
\Lambda_b (p,s)\rangle \nonumber \\ 
&=  \kappa \cdot \{ (C_7+C_7^\prime)
\langle \Lambda |\bar s \sigma_{\mu \nu} q^\mu b | \Lambda_b \rangle+
(C_7-C_7^\prime)
\langle \Lambda |\bar s \sigma_{\mu \nu} q^\mu \gamma_5 b | \Lambda_b \rangle  \} \epsilon^{* \nu} \, , 
\label{eq:Mfi}
\end{align}
where
\begin{equation}
\kappa \equiv -i\frac{G_F V_{tb} V_{ts}^* e m_b}{\sqrt{2}\, 4 \pi^2}
\end{equation}
and $p (s),p^\prime (s^\prime)$ are the four-momenta (spins) of the $\Lambda_b, \Lambda$, 
respectively, and $q,\epsilon$ denote the photon momentum and polarization.
In Eq.~(\ref{eq:Mfi}), $C_7, C_7^\prime$ (capital letters) are 
the effective coefficients of the corresponding dipole operators. 
At leading logarithmic order in $\alpha_s$, they coincide  with the Wilson 
coefficients $c_7, c_7^\prime$ (small letters), respectively, up to 
small contributions from penguin operators, and  
are understood to be evaluated at the $m_b$-scale. 

We write the hadronic matrix elements in terms of the 
tensors $\Gamma$ and $\Gamma_{(5)}$ defined as 
\begin{align}
\langle \Lambda |\bar s \sigma_{\mu \nu} q^\mu b | \Lambda_b \rangle & \equiv
\overline{\Psi}^\alpha (p^\prime, s^\prime)
\Gamma_{\alpha \nu} u(p^\prime+q,s) \, , \\
\langle \Lambda |\bar s \sigma_{\mu \nu} q^\mu \gamma_5 b| \Lambda_b \rangle 
 &\equiv \overline{\Psi}^\alpha (p^\prime, s^\prime) 
\Gamma_{(5) \, \alpha \nu} u(p^\prime+q,s) \, , 
\end{align}
where $\Psi^\alpha$ denotes the Rarita-Schwinger spinor of the $\Lambda(1520)$ 
and $u$ the $\Lambda_b$-spinor. 
To describe decays into on-shell photons, we use the following 
covariants with two
form factors $f_i \equiv  f_i(q^2=0)$, $i=1,2$:
\begin{align}
\label{eq:gamm}
\Gamma_{\alpha \nu} &= \frac{f_1}{m_{\Lambda_b}} 
(q_\alpha p^\prime_\nu - g_{\alpha \nu} p^\prime \cdot q) + 
f_2 (q_\alpha \gamma_\nu - g_{\alpha \nu} \not q) , \\
\Gamma_{(5) \, \alpha \nu} &=  \frac{f_1}{m_{\Lambda_b}} 
(q_\alpha p^\prime_\nu - g_{\alpha \nu} p^\prime \cdot q) \gamma_5 -
\left(\frac{m_\Lambda}{m_{\Lambda_b}} f_1 + f_2\right) 
(q_\alpha \gamma_\nu - g_{\alpha \nu} \not q) \gamma_5 .
\label{eq:gamm5}
\end{align}
Here, $m_{\Lambda_b}$, $m_\Lambda$ denote the masses of the $\Lambda_b$-
and $\Lambda$-baryon, respectively.
The vertices 
$\Gamma_{\alpha \nu}$ and $\Gamma_{(5) \, \alpha \nu}$ are related due to the identity 
$\sigma_{\mu \nu} \gamma_5=-\frac{i}{2}\epsilon_{\mu \nu \alpha \beta} \sigma^{\alpha \beta}$.
In writing down Eqs.~(\ref{eq:gamm}) and (\ref{eq:gamm5}),
we used the  equations of motion, 
gauge invariance $\Gamma_{\alpha \nu} q^\nu=0$ and
$\Gamma_{(5) \, \alpha \nu} q^\nu=0$, parity conservation of the strong interaction---for a 
$(1/2)^+ \to (3/2)^-$ transition  $\Gamma_{\alpha \nu}$ is P-even and $\Gamma_{(5) \, 
\alpha \nu}$ P-odd---and the conditions
\begin{equation}
\gamma_\alpha \Psi^{\alpha}=0 , \qquad   p^\prime_\alpha \Psi^\alpha =0 \, .
\end{equation}
For other Lorentz decompositions of $\Gamma,\Gamma_{(5)}$ we refer to Ref.~\cite{Jones-Scadron}.

We proceed to calculate the spin-averaged branching fraction for $ \Lambda_b \to \Lambda(1520) \gamma$ 
decays.
Employing the Rarita-Schwinger spin summation formula:
\begin{equation}
\sum_{s^\prime} \Psi_\alpha(p^\prime, s^\prime) \overline{\Psi}_\beta(p^\prime, s^\prime) = 
- (\not p^\prime + m_\Lambda) \left\{ g_{\alpha \beta} - \frac{\gamma_\alpha \gamma_\beta}{3} 
- \frac{2 p^\prime_\alpha p^\prime_\beta}{3 m_\Lambda^2} +\frac{p^\prime_\alpha \gamma_\beta
- p^\prime_\beta \gamma_\alpha}{ 3 m_\Lambda} \right\} \, ,
\end{equation}
we obtain for the total spin-averaged branching fraction with the 
$\Lambda_b$-lifetime $\tau_{\Lambda_b}$
\begin{equation}
\label{eq:br}
{\mathcal{B}}( \Lambda_b \to \Lambda(1520) \gamma)= \tau_{\Lambda_b} 
\left(1-\frac{m_\Lambda^2}{m_{\Lambda_b}^2}\right) 
\frac{ |{\mathcal{M}}_{\rm ave}|^2}{16 \pi m_{\Lambda_b}} \, , 
\end{equation}
where
\begin{align}
|{\mathcal{M}}_{\rm ave}|^2 =\ & \frac{1}{2} \sum_{\rm spins} |{\mathcal{M}}_{fi}|^2 \nonumber \\
=\ & \frac{2}{3} \frac{m_{\Lambda_b}^6}{m_\Lambda^2} 
\left( 1-\frac{m_\Lambda^2}{m_{\Lambda_b}^2} \right)^2 |\kappa|^2 (|C_7|^2+|C_7^\prime|^2) 
\left[ f_1^2 \left( \frac{m_\Lambda}{m_{\Lambda_b}}\right)^2 
\left(1 +\frac{m_\Lambda}{m_{\Lambda_b}}\right)^2 \right .\nonumber \\ 
 &+ \left. f_1 f_2 \frac{m_\Lambda}{m_{\Lambda_b}} 
\left(1+ 4  \frac{m_\Lambda}{m_{\Lambda_b}} +3  \frac{m_\Lambda^2 }{m_{\Lambda_b}^2}\right) 
+f_2^2 \left(1+ 3\frac{m_\Lambda^2}{m_{\Lambda_b}^2}\right) \right] .
\label{eq:mave}
\end{align}
In the limit $m_\Lambda \ll m_{\Lambda_b}$ this yields
\begin{equation}
\label{eq:brf2}
{\mathcal{B}}( \Lambda_b \to \Lambda(1520) \gamma)= \tau_{\Lambda_b} 
\left(1-\frac{m_\Lambda^2}{m_{\Lambda_b}^2} \right)^3 
\frac{m_{\Lambda_b}^5}{m_\Lambda^2} 
\frac{ \alpha G_F^2 |V_{tb} V_{ts}^*|^2 m_b^2}{192 \pi^4} 
f_2^2 (|C_7|^2+|C_7^\prime|^2)\, , 
\end{equation}
where we kept the full phase space factor.

There is no information on the form factors $f_{1,2}$ currently available.
In Sec.~\ref{sec:scet} we will work out a relation between them.
Since $|C_7|^2 + |C_7^\prime|^2$ is strongly constrained to be close to its 
SM value by data on $B \to X_s \gamma$ decays, a future 
measurement of the  $\Lambda_b \to \Lambda(1520) \gamma$ branching fraction
will determine $f_2$, see Eq.~(\ref{eq:brf2}).

Beyond lowest order, the relation between the genuine short-distance coefficients from
${\mathcal H}_{\mathrm{eff}}$, $c_7^{(\prime)}$, and the effective coefficients,
$C_7^{(\prime)}$, is modified by calculable
perturbative ${\mathcal{O}}(\alpha_s)$-corrections to the vertex, hard
scattering and
annihilation contributions, see, e.g., Ref.~\cite{Bosch-Buchalla} for the 
corresponding analysis of $B \to K^* \gamma $ decays, and also
Ref.~\cite{Ball-Jones-Zwicky} for contributions beyond QCD factorization.
Fully-fledged calculations for $b$-baryon decays are not available, but
we can still make some remarks:
The vertex correction at next-to-leading order
has already been estimated in Ref.~\cite{Hiller-Kagan} 
for $\Lambda_b \to \Lambda(1116) \gamma$ decays. 
We expect a similar correction for $\Lambda_b \to \Lambda(1520) \gamma$.
As for the contributions beyond the soft form factor, 
the hard-scattering scale for heavy-to-light baryons at large recoil
is lowered with respect to the one for mesons due to the larger number of
constituents, hence will induce parametrically larger 
$\alpha_s$-corrections. Weak annihilation contributions to
$b \to s$-transitions
are CKM-suppressed by $V_{ub} V_{us}^*/V_{tb} V_{ts}^*$, but arise at
tree level from $W$-boson exchange.
The situation regarding weak annihilation improves for baryons, 
where such contributions are colour-suppressed.

At higher order, the effective coefficients develop also an absorptive part, and
allow for CP violation in decay, see Ref.~\cite{Hiller-Kagan} for a discussion in
$\Lambda_b \to \Lambda(1116) \gamma$ decays. Similarly to final state
$\Lambda(1116)$'s, the heavier $\Lambda$ resonances decaying to $pK$ 
are self-tagging, i.e., $\Lambda$ decays to $K^-$ whereas  $\overline{\Lambda}$ 
decays to $K^+$.

%%%%%%%%%%%%%%%%%%%%%%%%%%%%%%%%%%%%%%%%%%%%%%%%%%%%%%%%%%%%%%%%%%%%%%%%%%%%%%%%%%%%%%%%%%%
\section{Helicity amplitudes for $\boldsymbol{\Lambda_b \to \Lambda(1520) \gamma}$ }
%%%%%%%%%%%%%%%%%%%%%%%%%%%%%%%%%%%%%%%%%%%%%%%%%%%%%%%%%%%%%%%%%%%%%%%%%%%%%%%%%%%%%%%%%%%

We decompose the decay amplitude into helicity amplitudes 
${\mathcal{A}}_h$, labelled by the $\Lambda$-helicity $h$, where $h=\pm 1/2, \pm 3/2$:
\begin{equation}
\label{eq:sum}
\sum_{\rm spins} | {\mathcal{M}}_{fi}|^2 =\sum_{h=\pm 1/2, \pm 3/2} | {\mathcal{A}}_h|^2 \, .
\end{equation}
We calculate the amplitudes $ {\mathcal{A}}_h$ in the $\Lambda_b$-rest frame, where we 
choose the momentum of the $\Lambda$ to be in the $+z$-direction, and use
\begin{gather}
p^{\prime \, \mu}=(E^\prime,0,0,E)\, , \qquad q^\mu=(E,0,0,-E) \, , \\
E=\frac{m_{\Lambda_b}^2-m_\Lambda^2}{2 m_{\Lambda_b}} \, , \qquad
E^\prime=\frac{m_{\Lambda_b}^2+m_\Lambda^2}{2 m_{\Lambda_b}} ,
\end{gather}
as well as $\epsilon^{* \, \mu}_{\pm 1}=\mp 1/\sqrt{2} (0,1,\mp i ,0)$ for a photon with 
angular momentum in the $+z$ direction, i.e., $J_z =\pm 1$ and 
$q \cdot \epsilon^*=0$. 
Note that in our reference frame we also have $p^\prime \cdot \epsilon^*_\pm=0$.
The $\Lambda$-polarization vectors $\omega_i$ with helicity $i = \pm 1,0$ are given as
\begin{equation} \label{eq:omega}
\omega^{* \, \mu}_{\pm 1} = \mp \frac{1}{\sqrt{2}} (0,1, \mp i,0) \, , \qquad
\omega^{* \, \mu}_{0} =\frac{1}{m_\Lambda} (E,0,0,E^\prime) \, , 
\end{equation}
with $p^\prime \cdot \omega_i^*=0$.
Then we write the Rarita-Schwinger spinor $\Psi^\alpha_h$ of the $\Lambda(1520)$ 
with helicity $h=\pm 1/2, \pm 3/2$ as
\begin{align}
\Psi_{\pm 3/2}^\alpha &= \Psi_{\pm 1/2} \cdot \omega^{\alpha}_{\pm 1} \, , \\
\Psi_{\pm 1/2}^\alpha &= \sqrt{\frac{2}{3}} \Psi_{\pm 1/2} \cdot \omega^{\alpha}_{0}+ 
\sqrt{\frac{1}{3}} \Psi_{\mp 1/2} \cdot  \omega^{\alpha}_{\pm 1} \, , 
\label{eq:psi12}
\end{align}
where in abuse of notation we denote by $\Psi_h$ the spin-1/2 component of the 
$\Lambda$ with helicity $h=\pm 1/2$.

The amplitudes ${\mathcal{A}}_{\pm 3/2}$ (${\mathcal{A}}_{\pm 1/2}$) result from a $\Lambda_b$-baryon 
with $h=\pm 1/2$ ($h=\mp 1/2$) and a photon with $J_z=\mp 1$, i.e., right- or left-handed,
respectively.
We arrive at the following helicity amplitudes for $ \Lambda_b \to \Lambda(1520) \gamma$ decays:
\begin{align}
\label{eq:hel1}
 {\mathcal{A}}_{+3/2}&= - 2\kappa \; (m_{\Lambda_b}^2 -m_\Lambda^2) \, 
\left[ \frac{m_{\Lambda_b}+m_\Lambda}{2 m_{\Lambda_b}} f_1 + f_2 \right] C_7^\prime \, , \\
 {\mathcal{A}}_{+1/2}&= - \frac{2\kappa}{\sqrt{3}} 
(m_{\Lambda_b}^2 -m_\Lambda^2) \left[ 
\frac{ m_{\Lambda_b}+ m_\Lambda}{2 m_{\Lambda_b}} f_1 +
\frac{ m_{\Lambda_b} }{m_\Lambda} f_2 \right] C_7^\prime \, , \\
{\mathcal{A}}_{-1/2}&= - \frac{2\kappa}{\sqrt{3}} (m_{\Lambda_b}^2 -m_\Lambda^2) 
\left[ \frac{m_{\Lambda_b}+ m_\Lambda}{2 m_{ \Lambda_b}} f_1 +
\frac{m_{\Lambda_b}}{m_\Lambda} f_2 \right] C_7 \, , \\
 {\mathcal{A}}_{-3/2}&= - 2\kappa \; (m_{\Lambda_b}^2 -m_\Lambda^2) \,
\left[ \frac{m_{\Lambda_b}+m_\Lambda}{2 m_{\Lambda_b}} f_1 + f_2 \right] C_7 \, .
\label{eq:hel4}
\end{align}
Evaluating Eq.~(\ref{eq:sum}), we recover the result for the squared 
spin-averaged matrix element given in Eq.~(\ref{eq:mave}).
The $1/m_\Lambda$ factor in ${\mathcal{A}}_{ \pm 1/2}$
results from the longitudinal polarization vector of the $\Lambda$-baryon, $\omega_0$.
It implies a (kinematical) suppression of the helicity-3/2 amplitude with respect to 
the helicity-1/2 one by
\begin{equation}
\label{eq:ratio}
  \frac{ {\mathcal{A}}_{ \pm 3/2} }{  {\mathcal{A}}_{ \pm 1/2} } \simeq \sqrt{3}
  \frac{m_\Lambda}{m_{\Lambda_b}} \left(1+ \frac{f_1}{2  f_2}\right)
\end{equation}
up to corrections of higher order in $m_\Lambda/m_{\Lambda_b}$.

%%%%%%%%%%%%%%%%%%%%%%%%%%%%%%%%%%%%%%%%%%%%%%%%%%%%%%%%%%%%%%%%%%%%%%%%%%%%%%%%%%%%%%%%%%%
\section{$\boldsymbol{\Lambda}$ helicity at large recoil and experimental
test}
%%%%%%%%%%%%%%%%%%%%%%%%%%%%%%%%%%%%%%%%%%%%%%%%%%%%%%%%%%%%%%%%%%%%%%%%%%%%%%%%%%%%%%%%%%%
\label{sec:scet}

Generally, we expect the  3/2-amplitude to be power-suppressed due the
enforced change
of helicity of the light degrees of freedom in the decay
\cite{Brodsky-Lepage,Burdman-Hiller}.
This can be shown explicitly using symmetry relations for
heavy-to-light currents arising when the emitted light hadron is energetic
\cite{CLOPR}.
In the symmetry limit, we obtain for $\Lambda_b \to \Lambda(1520)$
processes with
arbitrary Dirac structure
$\Gamma$:
\begin{align}
\langle \Lambda |\bar \chi \Gamma b_v | \Lambda_b \rangle &=
\overline{\Psi}^\alpha (n, s^\prime) q_{\alpha} X \Gamma u(v,s) \nonumber
\\
&= \overline{\Psi}^\alpha (n, s^\prime) q_{\alpha} (A + B \! \not v)
\Gamma u(v,s) \, ,
\end{align}
where $b_v$ is a heavy quark with velocity $v$ and $\chi$ a
strange-flavoured
collinear field associated with the light-like momentum
$n= p^\prime/E^\prime$, $n^2 \simeq 0$
living in the effective theory (SCET) \cite{BFPS}.
Wilson lines are understood in the definition of $\chi$.
In the second step, we have expressed the bispinor $X$ through a most
general, independent ansatz out of $v, n$ in terms of two form factors
$A,B$, both functions of $E^\prime$.
For radiative decays to on-shell photons with dipole currents
$\Gamma=\sigma_{\mu \nu} q^\mu (\gamma_5)$ this implies only one single
Dirac structure
\begin{equation}
\Gamma_{\alpha \nu}=\left(A + B \frac{m_\Lambda}{m_{\Lambda_b}}\right)
q_\alpha
\sigma_{\mu \nu} q^\mu
\end{equation}
and an analogous equation for $\Gamma_{(5)}$.
Comparing this to the full QCD formulae, Eqs.~(\ref{eq:gamm}) and
(\ref{eq:gamm5}), we obtain a relation between the form factors $f_1$ and
$f_2$:
\begin{equation} \label{eq:f1f2}
f_1 = - f_2 \frac{ 2 m_{\Lambda_b}}{ m_{\Lambda_b}+ m_{\Lambda}  } \, .
\end{equation}
Hence, at lowest order in SCET, the 3/2-amplitudes vanish, see
Eqs.~(\ref{eq:hel1}) and (\ref{eq:hel4}).
Similar to the analogous relations based on the helicity conservation
property
of the strong interaction in $B$ decays to vectors \cite{Burdman-Hiller,Hill}, 
Eq.~(\ref{eq:f1f2}) should hold to all orders in
$\alpha_s$.
We expect that finite ${\mathcal{A}}_{ \pm 3/2}$
arises from $1/E^\prime$-corrections to Eq.~(\ref{eq:f1f2}).

The form factor relation given in Eq.~(\ref{eq:f1f2}) is a central result
of our work.
It enables the extraction of the ratio
$|C_7^\prime/C_7|$ from $\Lambda_b$ decays to spin-3/2 $\Lambda$ baryons
as advocated in Ref.~\cite{Legger-Schietinger}, which requires disparate
helicity amplitudes.
Indeed, the squared amplitude ratio $\eta$, defined as
\begin{equation} \label{eq:eta}
\eta \equiv  \frac{ | {\mathcal{A}}_{ \pm 3/2}|^2 }{ | {\mathcal{A}}_{ \pm
1/2}|^2 }
\end{equation}
and introduced in Ref.~\cite{Legger-Schietinger}
to parameterize the suppression in sensitivity to $|C_7^\prime/C_7|$ from
a possible dilution
due to finite helicity-3/2 amplitude, turns out to be zero at lowest order
SCET: \begin{equation} \label{eq:eta0}
\eta = 0 \, .
\end{equation}
We expect finite $\eta$ at the order 
$(m_\Lambda/m_{\Lambda_b} \times \Lambda_\mathrm{QCD}/E^\prime)^2$
not exceeding the percent-level.

Interestingly the parameter $\eta$ can be determined experimentally from
\begin{equation}
\frac{d \Gamma}{d \cos \theta_p} \propto 1- \alpha_{p,3/2} \cos^2 \theta_p
\, , \quad \text{with} \quad
\alpha_{p,3/2} =\frac{\eta-1}{\eta+\frac{1}{3}} \, ,
\end{equation}
where $\theta_p$ denotes the angle between the $\Lambda$ direction and the
proton momentum in the $\Lambda$-rest frame \cite{Legger-Schietinger}.
We find that a sample of $10^4$
$\Lambda_b\rightarrow \Lambda(1520)\gamma\rightarrow pK\gamma$ decays,
equivalent to about three years of data-taking with the LHCb detector
\cite{LHCb-Reopt-TDR}, would yield a statistical precision of 0.03
on $\eta$. 
With a clear-cut prediction for $\eta$ at hand, this measurement would
provide a valuable benchmark for SCET.

%%%%%%%%%%%%%%%%%%%%%%%%%%%%%%%%%%%%%%%%%%%%%%%%%%%%%%%%%%%%%%%%%%%%%%%%%%%%%%%%%%%%%%%%%%%
\section{Photon helicity analysis}
%%%%%%%%%%%%%%%%%%%%%%%%%%%%%%%%%%%%%%%%%%%%%%%%%%%%%%%%%%%%%%%%%%%%%%%%%%%%%%%%%%%%%%%%%%%
\label{sec:photon}

\begin{center}
\begin{figure}[tb]
\hskip 1cm
      \includegraphics[width=12cm]{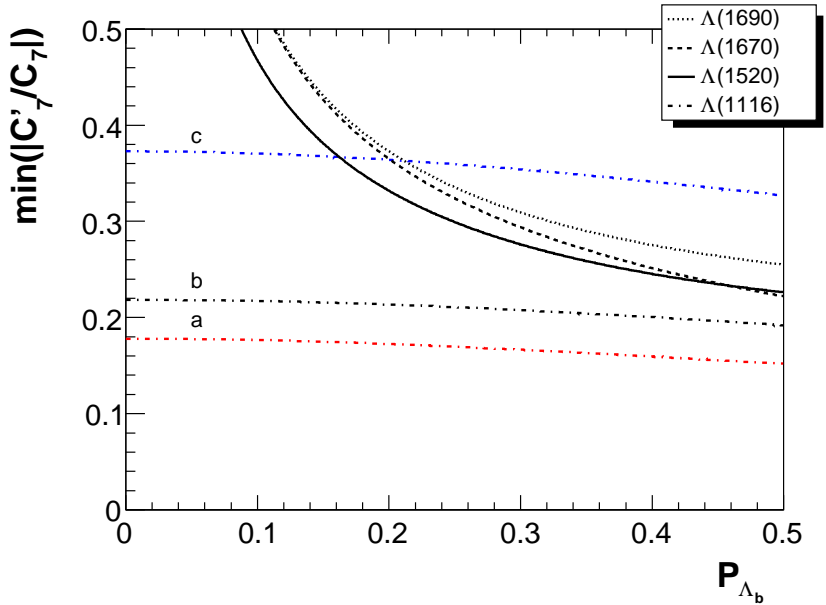}
\vskip -0.7 cm
  \caption{Experimental reach for the ratio 
    $|C_7^\prime/C_7|$ as a function of the $\Lambda_b$ polarization 
    $P_{\Lambda_b}$ for the decays 
    $\Lambda_b\rightarrow \Lambda(X)\gamma\rightarrow pK\gamma$, 
    $X=1520,1670,1690$.
    For comparison, the corresponding reach of the 
    $\Lambda_b\rightarrow \Lambda(1116)\gamma\rightarrow p\pi\gamma$
    decay is shown for the reconstruction scenarios (a)--(c) described in 
    the text.
    The curves indicate the minimum $|C_7^\prime/C_7|$ accessible at
    3$\sigma$ (standard deviation) significance at a hadron
    collider experiment after collecting statistics equivalent to $10^4$
    $\Lambda_b\rightarrow \Lambda(1520)\gamma\rightarrow pK\gamma$ decays.
    }
    \label{fig:sens}
\end{figure}
\end{center}

With known and finite $\Lambda_b$ polarization $P_{\Lambda_b}$ 
the helicity of the emitted photon, and hence the ratio $|C_7^\prime/C_7|$, is extracted 
from the differential $\Lambda_b \to \Lambda(1520) \gamma$ rate
(see Ref.~\cite{Legger-Schietinger} for details),
\begin{equation} \label{eq:dgammdtheta}
\frac{d \Gamma}{d \cos \theta_\gamma} \propto 1- \alpha_{\gamma,3/2} P_{\Lambda_b} 
\cos \theta_\gamma \, ,
\end{equation}
where the angle $\theta_\gamma$ between the $\Lambda_b$-spin and the direction of the photon momentum 
is defined in the $\Lambda_b$-rest frame.
For example, the SM-type amplitudes
${\mathcal{A}}_{-1/2} ({\mathcal{A}}_{-3/2})$ predominantly yield $\theta_\gamma \approx \pi (0)$.
The photon asymmetry parameter $\alpha_{\gamma,3/2}$ provides the link between experiment and
theory. 
It is given as
\begin{equation} \label{eq:alpha}
\alpha_{\gamma,3/2} =\frac{1-\eta}{1+\eta} \cdot 
\frac{|C_7|^2 -|C_7^\prime|^2}{|C_7|^2 +|C_7^\prime|^2} \, ,
\end{equation}
with $\eta$ defined in Eq.~(\ref{eq:eta}).
It is clear from Eq.~(\ref{eq:alpha}) that the sensitivity to $|C_7^\prime/C_7|$ would vanish
for $\eta \approx 1$.
Now, according to Sec.~\ref{sec:scet}
$\eta \approx 0$ and the extraction of $|C_7^\prime/C_7|$ from a measurement of 
$d \Gamma/d \cos \theta_\gamma$ is feasible without restriction. 
In Fig.~\ref{fig:sens} we show the expected experimental
sensitivity to right-handed currents in the case $\eta \approx 0$ ($f_1/f_2 \approx -2$)
as a function of the $\Lambda_b$ polarization $P_{\Lambda_b}$ for decays to 
$\Lambda(1520)$ and $\Lambda(1690)$.
The curves show the minimum $|C_7^\prime/C_7|$ accessible at 
3$\sigma$ (standard deviation) significance under the statistics assumptions of 
Ref.~\cite{Legger-Schietinger}, i.e., for a generic hadron collider 
experiment capable of collecting $10^4$
$\Lambda_b\rightarrow \Lambda(1520)\gamma\rightarrow pK\gamma$ decays.
The event yields underlying our sensitivity estimates are therefore simply
$10^4$ for $\Lambda(1520)$ and $10^4$ scaled by the expected branching
ratios \cite{Hiller-Kagan,Legger-Schietinger} for other $\Lambda$'s. 
The significance is determined as $(1-\alpha_{\gamma,\frac{3}{2}})/
\sigma_{\alpha_{\gamma,\frac{3}{2}}}$ (or equivalent for decays to spin-1/2 baryons),
see Erratum to Ref.~\cite{Legger-Schietinger}.

For comparison, the curves for decays to the spin-1/2 resonance $\Lambda(1670)$ and
to the ground state $\Lambda(1116)$, with $\Lambda(1116)\rightarrow p\pi$, are also
shown.
In the latter case, we illustrate the effect of event losses from trigger and
reconstruction difficulties by considering three scenarios in which the 
reconstruction efficiency for 
$\Lambda_b\rightarrow$ $\Lambda(1116)\gamma$ $\rightarrow p\pi\gamma$ is
(a) equal, 
(b) worse by factor of 10, and 
(c) worse by a factor of 100 with respect to 
$\Lambda_b\rightarrow \Lambda(X)\gamma\rightarrow pK\gamma$.
The expected reduction of the reconstruction efficiency is primarily caused
by the lack of an observable decay vertex in the innermost part of the detector.

\begin{figure}[tb]
\hskip 1cm
\vskip 1 cm
\end{figure}

Thanks to the suppression of the helicity-3/2 amplitude, radiative $\Lambda_b$ decays 
to spin-3/2 resonances suffer almost no loss in sensitivity to $|C_7^\prime/C_7|$
with respect to decays to spin-1/2 resonances for $\Lambda_b$ polarization not exceeding
about 0.2.
This is of vital importance since the cleanest and statistically most fertile among the 
$\Lambda_b\rightarrow pK\gamma$ resonance decays proceeds via the $\Lambda(1520)$, which
has spin 3/2.
For larger $\Lambda_b$ polarization, the statistical error on the proton asymmetry
parameter curbs the reach of decays to spin-3/2 resonances in comparison to decays 
to spin-1/2 resonances.
For instance for $P_{\Lambda_b} \approx 0.46$ and higher, the decay to $\Lambda(1670)$ 
can probe for lower $|C_7^\prime/C_7|$ than the decay to $\Lambda(1520)$, despite 
the inferior event yield of the $\Lambda(1670)$, see Fig.~\ref{fig:sens}.

%%%%%%%%%%%%%%%%%%%%%%%%%%%%%%%%%%%%%%%%%%%%%%%%%%%%%%%%%%%%%%%%%%%%%%%%%%%%%%%%%%%%%%%%%%%
\section{$\boldsymbol{\Lambda_b}$-polarization in hadronic collisions}
%%%%%%%%%%%%%%%%%%%%%%%%%%%%%%%%%%%%%%%%%%%%%%%%%%%%%%%%%%%%%%%%%%%%%%%%%%%%%%%%%%%%%%%%%%%
\label{sec:pol}

The extraction of the photon helicity from radiative $\Lambda_b$-decays
to $\Lambda$'s heavier than the 1116-MeV ground state
hinges crucially on the value of $\Lambda_b$-polarization $P_{\Lambda_b}$.
We seize this opportunity to briefly revisit the mechanism 
responsible for heavy baryon polarization at hadron colliders.

$\Lambda_b$-polarization is a consequence of finite $b$-quark polarization. The
latter arises in (unpolarized) $p \bar p$-, $pp$-collisions from QCD 
through its well-known mechanism of inducing CP-odd observables such 
as transversal spin asymmetries $\vec s \cdot (\vec n_1 \times \vec n_2)$
from strong phases.
We denote by $\sigma(\uparrow) (\sigma(\downarrow)$) the single quark 
production
cross sections with the quark spin $\vec s$ being up (down)
with respect to the scattering plane $\vec n_1 \times \vec n_2$.
Here, for example, the unit vectors $\vec n_{1}$ and $\vec n_{2}$ could be chosen 
along the beam direction and in the direction of the produced quark or hadron,
respectively.
This leads to the transversal quark-polarization 
\begin{equation}
  P_q \equiv \frac{\sigma (\uparrow) -\sigma (\downarrow)}{\sigma (\uparrow) 
 +\sigma (\downarrow)} \, .
\end{equation}
In high energy reactions, $P_q$ can be calculated in QCD-perturbation theory
\cite{Dharmaratna-Goldstein}.
Since it involves a quark helicity change, it is proportional to the 
quark mass.
$P_q$ depends non-trivially on the kinematics, such as the
centre-of-mass energy and scattering angle $\theta$. 
Note that for  $\theta \to 0 $ or  $\pi$, $P_q$ vanishes due to the absence 
of a scattering plane and emergent rotational invariance.
A finite $P_q$ arises at one-loop order in the strong interaction, hence
$P_q \propto \alpha_s m_q$. 
From explicit calculation the largest polarization for $b$-quarks is found to be
${\mathcal O}$(10\%) \cite{Dharmaratna-Goldstein}.
The dependence of $P_b$ on the kinematics could also result in different 
values of $P_{\Lambda_b}$ if measured in different kinematical regions or 
experimental set-ups. This deserves further study.

Finite polarization of $\Lambda_b$-baryons is then inherited from the
$b$-quarks as a fraction of the $b$-quark polarization.
To what extent depends on whether the $\Lambda_b$'s hadronize
directly from the $b$-quarks ($P_{\Lambda_b}=  P_b $ in the heavy quark
limit \cite{Mannel-Schuler}), or are produced via $\Sigma_b^{(*)}$-baryons,
which decay strongly to $\Lambda_b \pi$ \cite{Falk-Peskin}.
In the latter case, the amount of depolarization can be expressed in terms of
two fragmentation parameters in the limit of completely incoherently
decaying $\Sigma_b$, $\Sigma_b^*$-resonances. There is a lower bound
$P_{\Lambda_b}/P_b \geq 1/9$, but more common values for the parameters
yield much larger $P_{\Lambda_b}/P_b$, e.g., $72\%$ \cite{Falk-Peskin},
which is also supported by LEP data \cite{ALEPH-OPAL}. 
Besides fragmentation, the amount of depolarization also depends on how fast
the $(\Sigma_b, \Sigma_b^*)$-multiplet decays with respect to its energy 
splitting $\Delta$: first, the longer the $\Sigma_b^{(*)}$-resonances live, 
the more time there is for the heavy quark spin to interact with the light 
degrees of freedom and, second, the further apart the resonances are within 
the multiplet, the larger the heavy quark spin interaction.
Hence the $\Lambda_b$-polarization increases with
$\Gamma (\Sigma_b^{(*)} \to  \Lambda_b \pi)/\Delta$.
The dependence of the polarization on $\Gamma/\Delta$ has been
modelled by Falk and Peskin \cite{Falk-Peskin}.

Recently, the CDF collaboration reported a first measurement of
the mass splitting of the $\Sigma_b^{(*)}$-doublet \cite{CDF8523}
\begin{equation}
 \Delta \equiv m_{\Sigma^{* \mp}_b}- m_{\Sigma_b^\mp} =21.3^{+2.0
+0.4}_{-1.9-0.2} \,  \mbox{MeV}
 \, .
\end{equation}
In absence of a measurement the rate for $\Sigma_b^{(*)} \to \Lambda_b \pi$
decays can be calculated using heavy quark symmetries \cite{YK}
\begin{equation}
\Gamma \equiv
\Gamma (\Sigma_b^{(*)} \to  \Lambda_b \pi) =
\frac{1}{6 \pi} \frac{m_{\Lambda_b}}{m_{\Sigma_b^{(*)}}}
\frac{g_A^2}{f_\pi^2} | \vec p_\pi |^3 \, ,
\end{equation}
where $g_A = 0.75$ is a phenomenological coupling of the constituent quark, 
$f_\pi = 92.4$ MeV \cite{PDG06} is the pion decay constant, and
$\vec p_\pi$ denotes the pion momentum in the $\Sigma_b$ centre-of-mass
frame.
Note that the decay rates are equal up to higher order $1/m_b$ corrections.
Numerically the range 5808--5837 MeV for $\Sigma_b^{(*)}$ masses
\cite{CDF8523}
yields 6.5 MeV $< \Gamma <  14.7$ MeV, therefore
$0.3  < \Gamma/\Delta <  0.7$.
With rate and splitting being of the same order, the
$\Sigma_b,\Sigma_b^*$-resonances are partly overlapping, and the
depolarization
of the final $\Lambda_b$'s is reduced with respect to the
$\Gamma/\Delta \ll 1$ limit.

With this new information, we update the result from Ref.~\cite{Falk-Peskin}, 
using the same values for the fragmentation parameters, and find
\begin{equation}
74\% < \frac{P_{\Lambda_b}}{P_b} < 81\% \, .
\end{equation}
This range should be seen as a first order estimate;
it has uncertainties from the fragmentation parameters and the extrapolation 
to realistic values for $\Gamma$ and  $\Delta$.

We are aware that perturbative QCD is not sufficient to
explain the observed huge strange hyperon polarization \cite{ABDM}. 
Given the lack of a model-independent description, however,
which would then also apply to heavy baryons, 
we do not draw any conclusions for the polarization of the $\Lambda_b$'s.

Experimentally, $P_{\Lambda_b}$ is expected to be measured with a precision of
$\sigma(P_{\Lambda_b}) \approx 0.016$ at the LHC \cite{Smizanska}.

%%%%%%%%%%%%%%%%%%%%%%%%%%%%%%%%%%%%%%%%%%%%%%%%%%%%%%%%%%%%%%%%%%%%%%%%%%%%%%%%%%%%%%%%%%%
\section{Summary and Outlook}
%%%%%%%%%%%%%%%%%%%%%%%%%%%%%%%%%%%%%%%%%%%%%%%%%%%%%%%%%%%%%%%%%%%%%%%%%%%%%%%%%%%%%%%%%%%

Motivated by its prospective use as a probe for right-handed currents in
$b \to s \gamma$ we have worked out parameterizations for branching 
fraction and helicity amplitudes of the decay
$\Lambda_b \rightarrow \Lambda(1520)\gamma$.
Our framework applies to radiative $\Lambda_b$-decays to
any strange isosinglet baryon with $J^P=3/2^-$, that is, the
$\Lambda(1520)$, $\Lambda(1690)$
and so on.
Formulae for the even-parity $\Lambda$-baryons with $J^P=3/2^+$ can be obtained
by interchanging the right-hand side of Eq.~(\ref{eq:gamm}) with the one of
Eq.~(\ref{eq:gamm5}).
In the final formulae, Eqs.~(\ref{eq:br}--\ref{eq:brf2}) and
Eqs.~(\ref{eq:hel1}--\ref{eq:hel4}),
this corresponds to replacing $C_7^\prime$ by $-C_7^\prime$.

In the approximation of energetic light-hadron emission (SCET) we find the 
helicity-3/2 amplitude to vanish at lowest order, a result which simplifies
considerably the experimental extraction of the photon helicity in that mode,
as shown in Fig.~\ref{fig:sens}.
The predicted suppression of the helicity-3/2 amplitude is readily
testable at collider experiments and therefore could serve as a reference point 
for SCET and its applicability to baryons.
We have also re-estimated $\Lambda_b$-polarization at hadron colliders,
updating the prediction from heavy-quark effective theory with recent data on
$\Sigma_b^{(*)}$ resonances from the CDF collaboration.

Given sufficient data, the experimental investigations of radiative
$\Lambda_b$ decays can contribute further to the general programme of 
flavour physics \cite{SuperB}:
There is the possibility of searching for CP violation beyond the SM
by measuring the branching fractions and photon polarization
asymmetries $\alpha_{\gamma,3/2}$ in $\Lambda_b \to \Lambda(1520) \gamma$
and its CP conjugated mode  $\overline{\Lambda}_b \to \overline{\Lambda}(1520) \gamma$
separately. 
In the absence of new CP phases, the respective observables
from $b$ and $\bar b$ decay are equal up to a very small SM background of the
order ${\mathrm{Im}}(V_{us}^* V_{ub}/V_{ts}^* V_{tb})$.
Note that a non-zero CP asymmetry in both cases requires a finite strong phase,
which arises at higher order, see Sec.~\ref{sec:amplitude-and-bf}.
For the $\Lambda_b$--$\overline{\Lambda}_b$ rate asymmetry, new physics 
signals are already strongly constrained by $B \to K^* \gamma$ decays
\cite{Schietinger}, but in the presence of right-handed currents, the
asymmetry $\alpha_{\gamma,3/2}$ in baryon decays would provide
complementary information.
This has been worked out and discussed in detail for the case of the 
$\Lambda(1116)$ in Ref.~\cite{Hiller-Kagan} and can be applied to heavier 
$\Lambda$'s accordingly.

Another application for $\Lambda_b$ samples of very high statistics are
investigations of the $b \to d \gamma$ penguin with
$\Lambda_b \to N(X) \gamma \to p \pi \gamma$ decays.
Indeed, a primary goal of flavour physics is to find out whether the
Yukawa couplings are the only source of flavour and CP violation
or whether there are new such sources \cite{Hiller}.
The former case, which includes the SM, is termed \emph{Minimal Flavour
Violation} \cite{MFV} and subject to intense ongoing tests.
In this scenario, all flavour-changing processes are related via the
CKM-link, in particular the $b \to s \gamma$ and
$b \to d \gamma$ transitions.
We therefore also suggest the study of $\Lambda_b \to N(X) \gamma \to p
\pi \gamma$
decays in analogy to $\Lambda_b \to \Lambda(X) \gamma \to p K \gamma$,
that is, to measure branching fractions and photon polarization.
The $\Lambda_b \to N(X) \gamma$ modes offer one of the rare opportunities
to obtain information on right-handed currents in the $b \to d \gamma $
transition, another one being time-dependent CP asymmetries in $B$-meson decays to
CP eigenstates \cite{Atwood-Gronau-Soni}.
We expect to see the $b \to d \gamma$ transition in radiative
$\Lambda_b$ decays with branching fractions of a few $10^{-7}$--$10^{-6}$,
lowered by $|V_{td}/V_{ts}|^2 \approx 1/25$ with respect to the
corresponding $b \to s \gamma$ modes and in agreement with
$B \to (K^*,\rho) \gamma$ decays \cite{Schietinger}.
Currently, the $b \to d \gamma$ penguin is only poorly known,
and any bound on its chirality would add information.
The observation of a discrepancy between $C_7^\prime/C_7$
extracted from $b \to s $ versus $b\to d$ modes%
\footnote{In $b \to d \gamma$ decays,
the $C_7^{(\prime)}$ are understood to be effective coefficients defined
as in Eq.~(\ref{eq:Mfi}) with obvious flavour replacements and
${\mathcal{H}}_{\rm eff}$ having further
contributions from 4-quark operators $\propto V_{ub} V_{ud}^*$,
see, e.g., Ref.~\cite{Buchalla} for details.}
would be a clean signal of
breakdown of the SM and more generally Minimal Flavour Violation, where
at the perturbative level
\begin{equation} \label{eq:mfvtest}
\left| \frac{C_7^\prime}{C_7}  \right|_{b \to s} -
\left| \frac{C_7^\prime}{C_7}  \right|_{b \to d} =
 {\mathcal O}\left(\frac{m_s-m_d}{m_b}\right)
 \simeq {\mathcal O}\left(\frac{m_s}{m_b}\right) \, .
\end{equation}
Allowing also for non-perturbative effects, the difference
is still protected by U-spin symmetry, which is only mildly broken,
for example by the difference between baryon masses.
In addition, Eq.~(\ref{eq:mfvtest}) will receive corrections
%$\propto \frac{V_{ub} V_{ud}^*}{ V_{tb} V_{td}^*}
%\frac{\Lambda_{QCD}}{m_b}$
$\propto (V_{ub} V_{ud}^*/V_{tb} V_{td}^*)
(\Lambda_\mathrm{QCD}/{m_b})$
from annihilation and up-charm loops with tree-level induced Wilson
coefficient, which are not CKM suppressed in the $b \to d$ transition
\cite{GGLP}.
Studies in $B \to (K^*,\rho) \gamma$ decays show that these corrections
do not spoil the suppression of right-handed photons in $b \to (s,d) \gamma$
in the SM and give $|C_7^\prime/C_7|_{b \to (s,d)}$
of comparable order of magnitude \cite{GGLP,Ball-Jones-Zwicky}.

We would like to stress here that due to the interference
between mixing and decay, time-dependent CP asymmetries in meson decays
can only probe contributions with weak CP phase other than the one
of the meson mixing amplitude.
On the other hand, the photon polarization analysis in
$\Lambda_b \to \Lambda(X) \gamma, N(X) \gamma$
is sensitive to the magnitudes of the total amplitudes $|C_7^{(\prime)}|$.
We conclude that radiative $\Lambda_b$ decays constitute a rich testing
ground for physics within and beyond the Standard Model.

%%%%%%%%%%%%%%%%%%%%%%%%%%%%%%%%%%%%%%%%%%%%%%%%%%%%%%%%%%%%%%%%%%%%%%%%%%%%%%%%%%%%%%%%%%%
\section*{Acknowledgements}
%%%%%%%%%%%%%%%%%%%%%%%%%%%%%%%%%%%%%%%%%%%%%%%%%%%%%%%%%%%%%%%%%%%%%%%%%%%%%%%%%%%%%%%%%%%

The work of G.H.\ is supported in part by Bundesministerium f\"ur Bildung und Forschung, 
Berlin/Bonn.
T.S.\ gratefully acknowledges support from the Swiss National Science Foundation under 
grant Nr.~620-066162 during the early stages of this work.

%%%%%%%%%%%%%%%%%%%%%%%%%%%%%%%%%%%%%%%%%%%%%%%%%%%%%%%%%%%%%%%%%%%%%%%%%%%%%%%%%%%%%%%%%%%

\end{document}